\documentclass[journal]{IEEEtran}

\usepackage{graphicx,multicol}
\usepackage{psfrag,amsbsy,pstricks,amsmath,amsfonts,amssymb,cite}
\usepackage{algorithm,algpseudocode}
\usepackage{epstopdf}
\usepackage{mathtools}
\usepackage{color}
\usepackage{graphicx}

\usepackage{booktabs}

\newcommand{\s}{\boldsymbol{s}}

\newcommand{\tra}{^{\textrm{T}}}

\newcommand{\CN}{\mathcal{CN}}

\begin{document}
%
\title{A BP-MF-EP Based Iterative Receiver for Joint Phase Noise Estimation, Equalization and Decoding }

\author{ Wei~Wang, Zhongyong~Wang, Chuanzong~Zhang, Qinghua Guo,~Peng~Sun, and~Xingye~Wang
\thanks{This work was supported by the National Natural Science Foundation
of China  (NSFC 61571402, NSFC 61201251). }
\thanks{W. Wang, Z. Wang and P. Sun  are with the School of Information Engineering, Zhengzhou University, Zhengzhou 450001, China (e-mail: iewwang@zzu.edu.cn; iezywang@zzu.edu.cn; iepengsun@gmail.com).}
\thanks{C. Zhang is with the School of Information Engineering, Zhengzhou University, Zhengzhou 450001, China, and the Department of Electronic Systems, Aalborg University, Aalborg 9220, Denmark (e-mail: ieczzhang@gmail.com).}
\thanks{Q. Guo is with the School of Electrical, Computer and Telecommunications Engineering, University of Wollongong, Wollongong, NSW 2522, Australia, and also with the School of Electrical, Electronic and Computer Engineering, the University of Western Australia, Crawley, WA 6009, Australia (e-mail: qguo@uow.edu.au).}
\thanks{X. Wang is with the School of Information Engineering, North China University of Water Resources and Electric Power, Zhengzhou 450000, China (e-mail: wangxingye6507@sina.com).} 
}

\maketitle

\begin{abstract}
In this work, with combined belief propagation (BP), mean field (MF) and expectation propagation (EP),  an iterative receiver is designed for joint phase noise (PN) estimation, equalization and decoding in a coded communication system. The presence of the PN results in a nonlinear observation model. Conventionally,  the nonlinear model is directly linearized by using the first-order Taylor approximation, e.g., in the state-of-the-art soft-input extended Kalman smoothing approach (soft-in EKS). In this work, MF is used to handle the factor due to the nonlinear model, and a second-order Taylor approximation is used to achieve  Gaussian approximation to the MF messages, which is crucial to the low-complexity implementation of the receiver with BP and EP. It turns out that our approximation is more effective than the direct linearization in the soft-in EKS with similar complexity, leading to significant performance improvement as demonstrated by simulation results.    

\end{abstract}


\IEEEpeerreviewmaketitle

 \section{Introduction}\label{Sec:intro}

\IEEEPARstart{L}{ocal} oscillators, which provide a reference signal for time and frequency synchronization, are one of the key modules in a communication system. The instability of oscillators results in phase noise (PN), which may severely affect the system performance \cite{Khanzadi2014}.

Various Bayesian and non-Bayesian approaches have been proposed to solve the PN problem. Bhatti et al. modelled the PN  with a discrete cosine transform (DCT) expansion~\cite{Bhatti2009}, where the DCT coefficients can be easily estimated. However, the DCT method is a non-Bayesian one, and it does not make use of the time dependence of the PN process.  In Bayesian methods such as particle filter~\cite{Merli2008}, Tikhonov parametric estimation~\cite{Colavolpe2005}, and extended Kalman smoothing (EKS)~\cite{Nissila2009}, PN is modelled as a Wiener process. The particle filtering method~\cite{Merli2008} needs to sample  the posteriori probability density function (PDF) of  continuous-valued PN variables, where a larger number of particles  yields better performance at the cost of higher complexity. The Tikhonov parametrization  method~\cite{Colavolpe2005}  (or called a von Mises distribution \cite{Senst2011}) is an iterative method to deal with the presence of strong PN  for AWGN channels. The intractable integral operation associated with  continuous variables is circumvented by constraining the PDF to Tikhonov distribution. However, the work in \cite{Colavolpe2005} was focused on AWGN channel, and a straightforward extension to the inter-symbol interference (ISI) channel which is allowed by incorporating a MAP equalizer will lead to complexity growing exponentially with the channel memory length.
In the soft input EKS (Soft-in EKS) method\footnote{The EKS method in~\cite{Nissila2009} was proposed for AWGN channels. It can be extended to the case of ISI channels, e.g., by incorporating the BP-EP algorithm  \cite{SUN2015} to handle ISI channels.}  proposed in ~\cite{Nissila2009}, the nonlinear observation model is directly linearized by using the first order Taylor expansion. Soft-in EKS has been  used  in single-input single-output (SISO) and multiple-input multiple-output (MIMO) systems~\cite{Nissila2009,Mehrpouyan2012,Nasir2013,Krishnan2012}. 

Recently, the message passing techniques, such as belief propagation (BP) \cite{Kschischang2001} and variational message passing (VMP) \cite{Winn2005}, have been widely used for iterative receivers design. BP is effective for discrete probability models and linear Gaussian models. 
A BP-based equalizer proposed in \cite{Colavolpe2011} has a linear complexity, which is much lower than that of the equalizer in \cite{Koetter2004}. 
The VMP method, also referred as mean filed (MF), is especially suitable for handling variables with exponential distributions. Recently, a unified message passing framework was proposed in~\cite{Riegler2013}, where BP and MF are merged to keep the virtues of BP and MF while avoid their  drawbacks. It has been applied to joint channel estimation and decoding in orthogonal frequency division multiplexing (OFDM) system \cite{Badiu2012},\cite{Badiu2013} and single carrier frequency domain equalization (SC-FDE) system \cite{Zhang2015}. In addition, expectation propagation (EP)\cite{Hu2006}  has been used to achieve Gaussian approximation to non-Gaussian messages, and combined EP and BP has been applied to flat-fading or ISI channel equalization, e. g.,  in \cite{Qi2007,SUN2015} .  
 

 In this paper, with combined BP, MF and EP, we propose an iterative approach to joint PN estimation, equalization and decoding  for a coded system over ISI channels. BP and EP are used to deal with the linear model for PN process  and   modulation and coding, while   MF is used to handle the factor due to the nonlinear observation model. Furthermore, the non-Gaussian MF messages are  approximated to be Gaussian by using the second-order Taylor expansion, which enables  low-complexity implementation of the receiver with BP and EP. Our approximation is more effective than the direct linearization of the nonlinear model in the soft-in EKS \cite{Nissila2009}, which is demonstrated by the significant performance gain of the proposed approach in terms of mean-square-error (MSE) of PN estimation and system bit-error-rate (BER) performance.  

\textit{Notation}-The superscriptions $(\cdot)\tra $and $(\cdot)^\text{H} $denote the transpose and conjugate transpose, respectively. We use $\propto $ to denote equality of functions up to a scale factor, and use $\boldsymbol I_N $  to denote an $ N\times N $ identity matrix. The real part of a complex quantity is denoted by $\Re[\cdot] $. The functions $\mathcal N(x;\hat{x},\sigma^2_x) $  and $\mathcal {CN}(x;\hat{x},\sigma^2_x) $ stand for real and proper complex  Gaussian probability distributions with mean $\hat{x} $ and variance $\sigma^2_x $, respectively.

\section{System Model and Factor Graph Representation}\label{Sec:SysMod}

We consider a coded communication system. An information bit sequence $ \boldsymbol b=[b_0,...,b_{N_b-1}]\tra $  is encoded and interleaved, yielding an interleaved codeword $\boldsymbol c=[c_0,...,c_{N_c-1}]\tra $. Then sequence $ \boldsymbol c$ is mapped to a symbol sequence $\boldsymbol x=[x_0,...,x_{M-1}]\tra $ which is transmitted over an ISI channel with coefficients $\boldsymbol h=[h_{L-1},...,h_{0}]\tra $.
The channel coefficients are assumed to be constant during each transmitted block and they are available to the receiver. By considering the effect of PN, the  received baseband signal at time instant $k$, $(k=0, 1, ..., M+L-2)$, can be represented as\cite{Nissila2006}
\begin{eqnarray}
y_k=e^{j\theta_k} \sum_{l=0}^{L-1}h_lx_{k-l}+n_k=e^{j\theta_k} \boldsymbol h^ {\textrm{T}}\boldsymbol s_k+n_k
\label{eq:sys1}
\end{eqnarray}
where  $\boldsymbol s_k\triangleq [x_{k-L+1},...,x_k]\tra $ with $x_k=0 $  for $k<0$ and $k>M-1$,   
and  $n_k $ is a sample of the complex Gaussian noise with variance $\sigma_n^2 $. The phase  $\theta_k$ represents the PN at time instant $k$,  and the PN  can be modelled as a random-walk (Wiener) process \cite{Colavolpe2005}, \cite{Nissila2009}
\begin{eqnarray}
\theta_k=\theta_{k-1}+\Delta \theta_k
\label{eq:winner}
\end{eqnarray}
where $\Delta \theta_k $ is a white real Gaussian process with distribution $\mathcal N(\Delta\theta_k;0,\sigma^2_{\Delta}) $, and  $\theta_0 $ is assumed to have a uniform distribution over $[0,2\pi) $.
 We define $\boldsymbol \theta =[\theta_0,\theta_1,...,\theta_{M+L-2}]\tra$.


The joint probability of $\boldsymbol{b,c,x,s}$ and $\boldsymbol{\theta}$ with given observation $\boldsymbol{y}=[y_0,y_1,\dots,y_{M+L-2}]\tra$ can be expressed as
\begin{align}
 p(\boldsymbol b, \boldsymbol c,\boldsymbol x,\boldsymbol s,\boldsymbol \theta|\boldsymbol y)\propto & \prod_{k=0}^{M+L-2} f_{y_k}(\boldsymbol s_k,\theta_k) f_{\boldsymbol s_k}(\boldsymbol s_k,\boldsymbol s_{k-1},x_k) \nonumber\\
& f_{\theta_0}\prod_{k=1}^{M+L-2} f_{\theta_k}(\theta_k,\theta_{k-1}) f_{\textrm{X}}(\boldsymbol x,\boldsymbol c,\boldsymbol b)
\label{eq:pdf}
\end{align}
where  $f_{y_k}(\boldsymbol s_k,\theta_k )\triangleq p(y_k|\boldsymbol s_k ,\theta_k)\propto \mathcal{CN}(y_k;e^{j\theta_k}\boldsymbol h^ {\textrm{T}}\boldsymbol s_k,\sigma_n^2 )$ denotes the likelihood function of  $\boldsymbol s_k $ and $\boldsymbol \theta_k $,   
$f_{\theta_k}(\theta_k,\theta_{k-1})\triangleq p(\theta_k|\theta_{k-1})
= \mathcal{N}(\theta_k;\theta_{k-1}, \sigma^2_{\Delta})$ is the conditional PDF of $\theta_k$ given $\theta_{k-1} $, and $f_\textrm{X}(\boldsymbol x,\boldsymbol c,\boldsymbol b)$ denotes the mapping, interleaving and coding constraints. 
Function $f_{s_k}\left(\boldsymbol s_k,\boldsymbol s_{k-1},x_k \right)$  represents the deterministic relationship between $\boldsymbol s_k $, $\boldsymbol s_{k-1} $ and $x_k $ which is given by $\boldsymbol s_k=\boldsymbol G\boldsymbol s_{k-1}+\boldsymbol e x_k $, where the $L\times L $ matrix $\boldsymbol G=[\boldsymbol 0 \ \boldsymbol I_{L-1};\ 0\ \boldsymbol 0\tra] $,  the length-$L $  vector $\boldsymbol e =[\boldsymbol 0\tra\ 1]\tra $, and  $\boldsymbol 0 $ is a zero column vector with length $L-1 $. 

\begin{figure}[!t]
\centering
\includegraphics[width=0.45\textwidth]{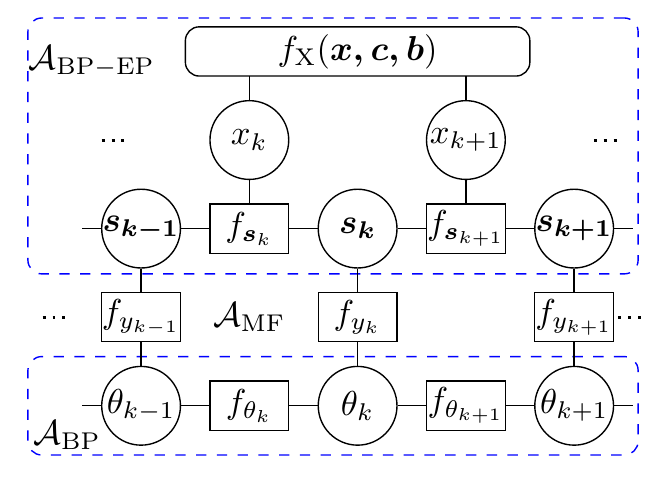}
\caption{Factor-graph representation of the probabilistic model~(\ref{eq:pdf})}
\label{fig:FG}
\end{figure}

A factor graph representation of \eqref{eq:pdf} is shown in Fig.~\ref{fig:FG}, which will be employed to develop a combined BP-MF-EP based receiver to achieve joint PN estimation, equalization and decoding in next section.

\section{Iterative Receiver Design with BP-MF-EP}\label{Sec:VMPBP}
Due to the presence of PN, the observation model in (\ref{eq:sys1}) is nonlinear. In EKS, the nonlinear model is directly linearized with the first order Taylor approximation.  The nonlinear model is represented by  the factors {$\{f_{y_k},\forall k\}$}  in Fig.~\ref{fig:FG}. In this work, we use MF to handle the factors.

As shown in Fig.~\ref{fig:FG}, we partition the graph into three parts: BP-EP subgraph, MF subgraph and BP subgraph. Accordingly, the factor nodes are classified into three disjoint sets: $\mathcal A_{\textrm{BP-EP}}\triangleq \{f_{s_k},f_{\textrm{X}}; \forall k\}$, $\mathcal A_{\textrm{MF}}\triangleq \{f_{y_k}; \forall k\}$ and $\mathcal A_{\textrm{BP}}\triangleq \{f_{\theta_k}; \forall k\}$    with $ \mathcal A_{\textrm{BP-EP}}\bigcap  \mathcal A_{\textrm{MF}}\bigcap  \mathcal A_{\textrm{BP}}=\emptyset$.  In the following, we detail the messages updating in each subgraph.


\subsection{ Message Passing in BP Subgraph}
As shown in Fig.~\ref{fig:FG}, message passing for PN estimation operates in  the BP subgraph, where we need to calculate the forward and backward messages and the outgoing messages which are input to the MF subgraph.

We assume that the incoming messages from the MF subgraph are available, and they are Gaussian, i.e., we have $\{m_{f_{y_k}\to \theta_k} \left(\theta_k \right)\propto \mathcal{N}( \theta_k;\hat{\theta}^\downarrow_{k}, \sigma^2_{\theta^\downarrow_k}),\forall k\}$.
The details on the calculations of the incoming messages are delayed to Section \ref{Sec:fy}. It is worth mentioning that, with the incoming Gaussian messages, all the messages running in the subgraph are Gaussian.

With the Gaussian message $m_{f_{\theta_{k-1}}\to\theta_{k-1 }}(\theta_{k-1})$ $\propto \mathcal{N}(\theta_{k-1};\hat\theta^\rightarrow_{k-1},\sigma^2_{\theta_{k-1}^\rightarrow})$, the message from variable $\theta_{k-1}$ to factor  $f_{\theta_k}$ is calculated as  $ n_{\theta_{k-1}\to f_{\theta_k}}(\theta_{k-1})= m_{f_{\theta_{k-1}}\to\theta_{k-1 }}(\theta_{k-1})m_{f_{y_{k-1}}\to\theta_{k-1} }(\theta_{k-1})$. The forward message $m_{f_{\theta_k}\to\theta_k }(\theta_k)$ reads
 \begin{align}\label{eq:forward}
m_{f_{\theta_k}\to\theta_k }(\theta_k)&\propto \int f_{\theta_k}(\theta_k,\theta_{k-1})n_{\theta_{k-1}\to f_{\theta_k}}(\theta_{k-1})d\theta_{k-1}\nonumber\\
&\propto\mathcal  N(\theta_k;\hat\theta^\rightarrow_k,\sigma^2_{\theta_k^\rightarrow}),
\end{align}
We assume that the initial phase noise $\theta_0$ is absorbed into the channel in the acquisition of the channel state information \cite{Mehrpouyan2012}, so the initial message for the forward recursive process $\hat\theta_0=0, \sigma^2_{\theta_0}=0$. 

Same to the forward messages, the backward message $m_{f_{\theta_{k+1}}\to\theta_k }(\theta_k) \propto\mathcal  N(\theta_k;\hat\theta^\leftarrow_k,\sigma^2_{\theta_k^\leftarrow})$.


According to \cite{Riegler2013}, the outgoing messages input to the MF subgraph should be the belief of $\theta_k$, which can be calculated as

\begin{align}\label{eq:saittofy}
b(\theta_k)&=m_{f_{y_k}\to \theta_k}(\theta_k)m_{f_{\theta_k}\to \theta_k}(\theta_k)
m_{f_{\theta_{k+1}}\to \theta_k}(\theta_k)\nonumber\\
&\propto \mathcal N(\theta_k;\hat\theta_k,\sigma^2_{\theta_k}),
\end{align}
where
\begin{align}
&\sigma^{-2}_{\theta_k}=\sigma^{-2}_{\theta_k^\downarrow}+\sigma^{-2}_{\theta_k^\leftarrow}+\sigma^{-2}_{\theta_k^\rightarrow}\\
&\hat\theta_k=\sigma^2_{\theta_k}(\sigma^{-2}_{\theta_{k}^\downarrow}\hat\theta_{k}^\downarrow+\sigma^{-2}_{\theta_{k}^\leftarrow}\hat\theta_{k}^\leftarrow+\sigma^{-2}_{\theta_{k}^\rightarrow}\hat\theta_{k}^\rightarrow) \label{eq:belief_theta}.
\end{align}

\subsection{Message Passing in BP-EP Subgraph}
We assume that the incoming messages from the MF subgraph are available, and they are  Gaussian. The calculations of the incoming messages will be detailed in Section \ref{Sec:fy}. 
So this subgraph involves the incoming Gaussian messages from the MF subgraph and discrete binary messages from the decoder. For this subgraph, we simply borrow the BP-EP algorithm developed in~\cite{SUN2015} where the use of EP produces Gaussian messages for ${x_k}$, which will in turn lead to Gaussian output messages in the BP-EP subgraph.  We refer readers to ~\cite{SUN2015} for the details of the BP-EP algorithm.    

With the BP-EP algorithm, we can calculate the  messages
 $ m_{f_{\boldsymbol s_k} \rightarrow \boldsymbol s_k   }(\boldsymbol s_k)\propto \CN (s_k;\hat{\boldsymbol s}^\rightarrow_k,\Sigma_ {\boldsymbol s^\rightarrow_k})$ 
 and $m_{f_{\boldsymbol s_{k+1}} \rightarrow \boldsymbol s_{k}   }(\boldsymbol s_{k}) \propto (s_k;\hat{\boldsymbol s}^\leftarrow_k,\Sigma_ {\boldsymbol s^\leftarrow_k}) $, which are all Gaussian.

According to \cite{Riegler2013}, the outgoing messages are the belief of $\s_k$ denoted by $b(\boldsymbol s_k)$, which are Gaussian again and can be expressed as

\begin{align}\label{eq:sktofyk}
n_{\boldsymbol s_k\rightarrow f_{y_k}}(\boldsymbol s_k) &\propto m_{f_{y_k} \rightarrow \boldsymbol s_k   }(\boldsymbol s_k)
m_{f_{\boldsymbol s_k} \rightarrow \boldsymbol s_k   }(\boldsymbol s_k) m_{f_{\boldsymbol s_{k+1}} \rightarrow \boldsymbol s_k   }(\boldsymbol s_k)\nonumber\\
&\propto\exp\left\{-(\boldsymbol s_k-\hat{\boldsymbol s}_k)^{\textrm {H}} \Sigma^{-1}_ {\hat{\boldsymbol s}_k}(\boldsymbol s_k-\hat {\boldsymbol s}_k)\right\}
\end {align}
where $m_{f_{y_k} \rightarrow \boldsymbol s_k   }(\boldsymbol s_k)$ is obtained by \eqref{eq:fyk} and
\begin{align}
&\Sigma^{-1}_ {\boldsymbol s_k}=\Sigma^{-1}_ {\boldsymbol s^\rightarrow_k}+\Sigma^{-1}_ {\boldsymbol s^\uparrow_k}+
 \Sigma^{-1}_ {\boldsymbol s^\leftarrow_k}
 \label{eq:svar1}\\
&\Sigma^{-1}_ {\boldsymbol s_k} \hat{\boldsymbol s}_k=\Sigma^{-1}_ {\boldsymbol s^\rightarrow_k}\hat{\boldsymbol s}^\rightarrow_k+\Sigma^{-1}_ {\boldsymbol s^\uparrow_k}\hat{\boldsymbol s}^\uparrow_k+
 \Sigma^{-1}_ {\boldsymbol s^\leftarrow_k}\hat{\boldsymbol s}^\leftarrow_k.
 \label{eq:esvar1}
\end{align}

\subsection{Message Passing in the MF Subgraph}\label{Sec:fy}

As shown by the middle part of the graph in Fig.~\ref{fig:FG}, the MF subgraph consists of the observation factors  $f_{y_k} $. We need to compute the outgoing messages to the BP-EP subgraph (BP subgraph)  based on the incoming messages from the BP subgraph (BP-EP subgraph). 

Assume that the incoming message  $b(\boldsymbol s_k)$ from  the BP-EP subgraph  is  available. According to the rules~\cite{Riegler2013} the outgoing messages to the BP subgraph can be computed as
\begin{align}
m_{f_{y_k}\rightarrow \theta_k}(\theta_k)&\propto\exp\left\{\int \log(f_{y_k}(\theta_k,\boldsymbol s_k))
    b(\s_k)d\boldsymbol s_k\right\}\nonumber\\
&\propto {\exp\left\{\Re [r_ke^{j\theta_k}]\right\}}
\label{eq:mfytheta1}
\end{align}
where {$r_k\triangleq 2\sigma^{-2}_ny_k^* \boldsymbol h^{\textrm T }\hat {\boldsymbol s}_k$}, and $\hat\s_k$ is the mean parameter vector of the Gaussian belief $b(\s_k)$. Note that the message $m_{f_{y_k}\rightarrow \theta_k}(\theta_k)$ yielded in \eqref{eq:mfytheta1} is no longer Gaussian. However, Gaussian messages  are expected for the BP subgraph for PN estimation, which is crucial to its low complexity implementation. To achieve this, we use the second-order Taylor expansion of $\Re [r_ke^{j\theta_k}]$ at the estimate of ${\theta_k}$, \text{i.e.},   
{\begin{align}
& m_{f_{y_k}\rightarrow \theta_k}(\theta_k) \nonumber\\ 
&\approx\exp\left\{ -\frac{1}{2}\Re[r_ke^{j\hat{\theta}_k}] \theta_k^2
+\Re[r_ke^{j\hat{\theta}_k}(j+\hat\theta_k)]\theta_k\right\} \nonumber\\
&\propto \mathcal N(\theta_k;\hat\theta_k^\downarrow,\sigma^2_{\theta_k^\downarrow})
\label{eq:fy_seita}
\end{align}}
where {$\hat{\theta}_k$ denotes the mean of $\theta_k$ computed in \eqref{eq:belief_theta}, and 
\begin{align}
&\sigma^{-2}_{\theta_k^\downarrow}=\Re [r_ke^{j\hat{\theta}_k}] 
\label{eq:var1}\\
&\sigma^{-2}_{\theta_k^\downarrow}\hat\theta^\downarrow_k=\Re [r_ke^{j\hat{\theta}_k}(j+\hat\theta_k)].
\label{eq:evar1}
\end{align}}

It is noted that the Soft-in-EKS algorithm~\cite{Nissila2009} uses the first-order Taylor expansion to locally linearize model~(\ref{eq:sys1}) directly. In contrast, we  use the second order Taylor series to approximate the MF message $ m_{f_{y_k}\rightarrow \theta_k}(\theta_k)$. It turns out that the performance of our algorithm is much better than that of the Soft-in-EKS algorithm, as demonstrated by simulation results.

Similarly, we also apply MF message update rules to the computation of  the outgoing message $m_{f_{y_k} \rightarrow \boldsymbol s_k   }(\boldsymbol s_k)$ for the BP-EP subgraph 
\begin{align}
 m_{f_{y_k} \to \boldsymbol s_k   }(\boldsymbol s_k)& =\exp\left\{\int\log(f_{y_k}(\theta_k,\boldsymbol s_k)) b( \theta_k)d\theta_k\right\}\nonumber\\
&\propto\exp\left\{-\boldsymbol s_k^{\textrm H}\Sigma^{-1}_ {\boldsymbol s^\uparrow_k}\boldsymbol s_k + 2\Re[\s_k^{\textrm H}\Sigma_{\boldsymbol s^\uparrow_k}^{-1}\hat{\boldsymbol s}^\uparrow_k] \right\}
 \label{eq:fyk}
\end {align}
where
\begin{align}
&\Sigma^{-1}_ {\hat{\boldsymbol s}^\uparrow_k}=\sigma_n^{-2} \boldsymbol h {\boldsymbol h}^{\textrm T}
\label{eq:hvar1}\\
&\Sigma_{\boldsymbol s^\uparrow_k}^{-1}\hat{\boldsymbol s}^\uparrow_k=\sigma_n^{-2}\Re[\ y_k\langle e^{-j\theta_k}\rangle_{b(\theta_k)}\boldsymbol h].
\label{eq:ehvar1}
\end{align}
An approximation to the term $\langle e^{-j\theta_k}\rangle_{b(\theta_k)}$ in~\eqref{eq:ehvar1} can be obtained by exploiting the second-order Taylor expansion, and it can be calculated as  
$\langle e^{-j\theta_k}\rangle_{b(\theta_k)}\approx e^{-j\hat{\theta}_k}(1-0.5\sigma^2_{\theta_k})$.
\subsection{Message Passing Scheduling }
 The overall message passing schedule for joint PN estimation, equalization and decoding is summarized in \textbf{Algorithm~\ref{alg:ww}}.


\begin{algorithm}
\caption{The combined BP-MF-EP Algorithm}\label{alg:ww}
\begin{algorithmic}[1]\small
\State\textbf{input} $\boldsymbol y,\boldsymbol h, \hat\theta_0,\sigma^2_{\theta_0} $
\State \textbf{initialize} $n_{\theta_0\to f_{\theta_1}}(\theta_0)$, $ m_{f_{y_k}\rightarrow {\theta_k}}(\theta_k) ,\forall k $ 
\For{$i=1\to$Iteration}
\State\textbf{for} $k=1\rightarrow M+L-2$, compute   $m_{f_{\theta_k}\rightarrow \theta_k}(\theta_k) $ using~(\ref{eq:forward})
\State\textbf{for} $k=M+L-3\rightarrow 1$, compute  $m_{f_{\theta_{k+1}}\rightarrow \theta_k}(\theta_k)$
\State\textbf{for all} $k$: compute $n_{\theta_k\to f_{y_k}}(\theta_k) $ using~(\ref{eq:saittofy})
\State\textbf{for all} $k$: compute $m_{f_{y_k}\rightarrow {\boldsymbol s_k}}(\boldsymbol s_k) $ using~(\ref{eq:fyk})
\State Run the BP-EP algorithm ~\cite{SUN2015} in the BP-EP subgraph
\State{\textbf{for all} $k$:}~update $n_{\boldsymbol s_k\to f_{y_k}}(\boldsymbol s_k) $ using~(\ref{eq:sktofyk})  
\State{\textbf{for all} $k$:}~update $ m_{f_{y_k}\rightarrow {\theta_k}}(\theta_k) $  using~(\ref{eq:fy_seita})  
\EndFor\ $i$
\end{algorithmic}
\end{algorithm}
\subsection{Complexity Analysis}
Note that the BP-EP algorithm in \cite{SUN2015} is incorporated in both the Soft-in EKS method and the proposed BP-MF-EP method to handle ISI channels. Hence, both methods involve the computation of (\ref{eq:sktofyk}), which requires a complexity of $O(L^3)$. We can also see that PN estimation in both the Soft-in EKS method and the BP-MF-EP method (i.e., the computation of (\ref{eq:fy_seita}) and the message passing in the BP subgraph shown in Fig.\ref{fig:FG}) have similar complexity, which is in the order of $L$. From the above analysis, the BP-MF-EP method and the Soft-in EKS method have similar complexity.
\section{SIMULATION RESULTS}
In this section, we evaluate the performance of the proposed method and compare it with the Soft-in EKS method (where the BP-EP algorithm in \cite{SUN2015} is incorporated to handle ISI channels) in terms of MSE for PN estimation and BER for the system performance. The system settings are as follows. The length of symbols in each frame is $1024 $. A rate-1/2 nonsystematic convolutional code with generator $(23,35)_8 $ is used to encode the bits sequence, and the coded sequence is permuted with a pseudo random interleaver. QPSK with Gray mapping is used. In simulations, the phase noise is generated using a Wiener process~(\ref{eq:winner}) with innovation variance $\sigma^2_\Delta=1\times 10^{-4 }$ and the Proakis-C channel with coefficients $\boldsymbol h=[0.227,0.460,0.668,0.460,0.227]\tra $ is used to examine the performance of the receiver. As in \cite{Covalope2005_2}, 5 pilot symbols are inserted every 256 symbols to make the iterative process bootstrap.


\begin{figure}[!t]
\centering
\includegraphics[width=0.5\textwidth]{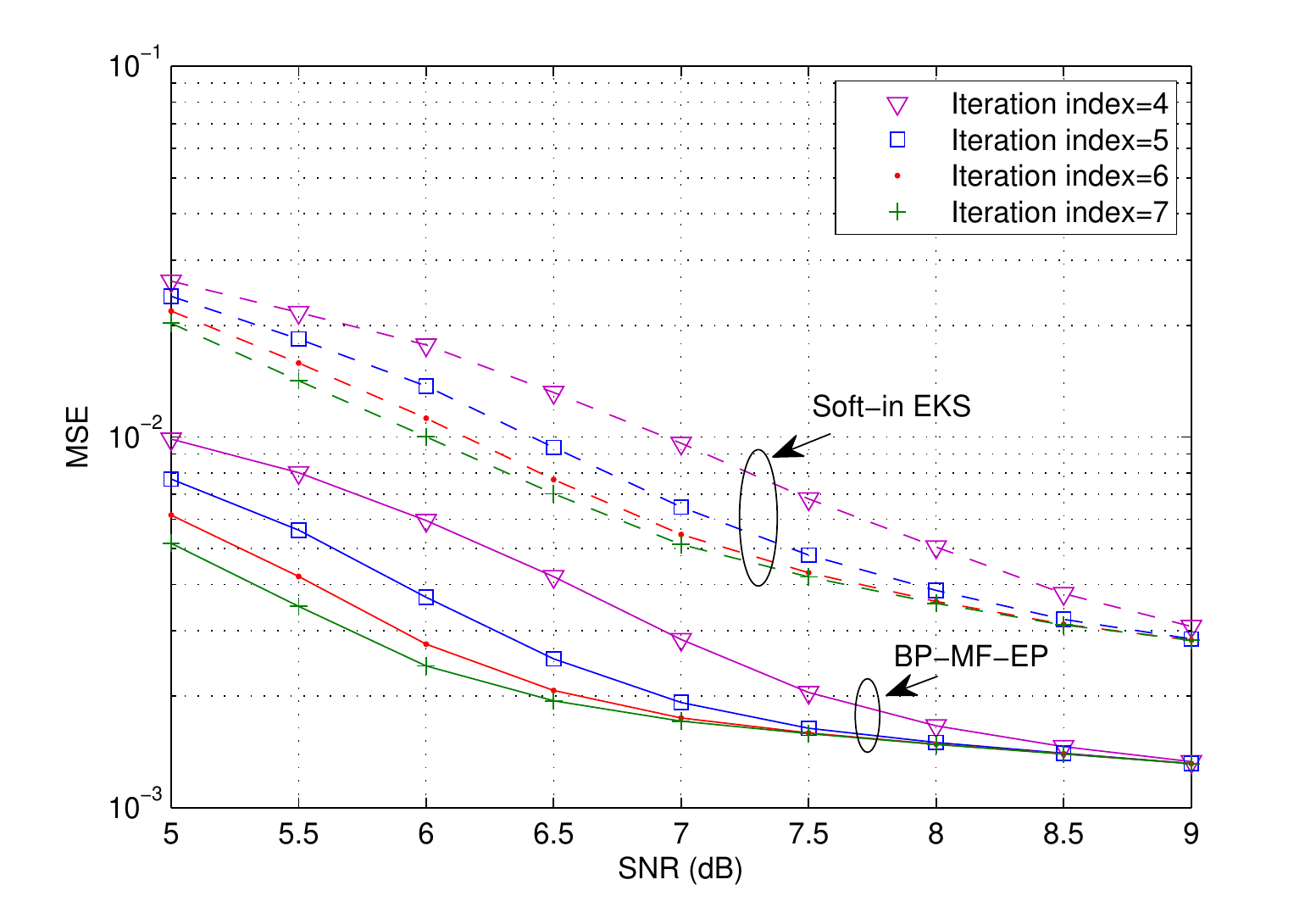}
\caption{  MSE performance of the phase noise estimation. }
\label{fig:MSE}
\end{figure}
\begin{figure}[!t]
\centering
\includegraphics[width=0.5\textwidth]{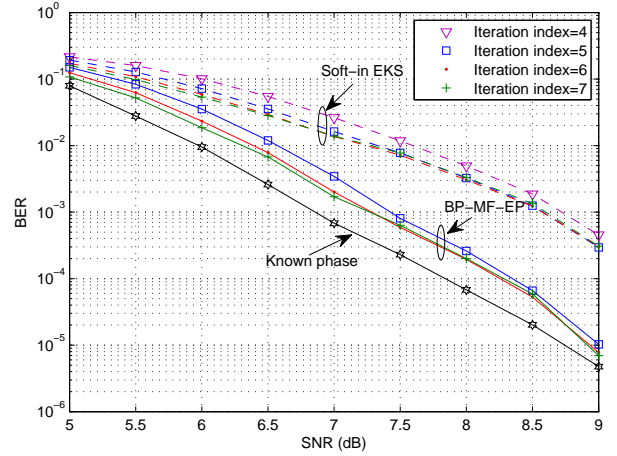}
\caption{  BER performance versus SNR.}
\label{fig:ber}
\end{figure}

We compare the MSE performance of the proposed algorithm with  that of the Soft-in EKS algorithm for PN estimation. The results with different number of iterations are shown in Fig.~\ref{fig:MSE}. It can be seen that the proposed BP-MF-EP method significantly outperforms the Soft-in EKS method.

The comparisons of system BER performance are shown in Fig.~\ref{fig:ber}, where the performance with known PN is also included for reference. It can be seen that considerable performance gains can be achieved by the proposed BP-MF-EP method compared to the Soft-in EKS method.


\section{CONCLUSIONS }
In this paper we have proposed an iterative receiver for joint PN estimation, equalization and decoding based on combined BP, MF and EP.  In particular, MF is used to tackle the factors due to the nonlinear observation model  and the second-order Taylor expansion is used to achieve Gaussian approximation to the MF messages, which is crucial to the low complexity implementation of the receiver. The approximation is more effective than the direct local linearization of the observation model in the Soft-in EKS. As shown by the simulation results,  the proposed method significantly outperforms the Soft-in EKS  with similar complexity.

%

\end{document}